%% file: JSAE26_PhysicsAI.tex
\documentclass[a4paper,10pt,twocolumn]{article}

\usepackage[T1]{fontenc}
\usepackage[utf8]{inputenc}
\usepackage{newtxtext,newtxmath}  

\usepackage[
  a4paper,
  top=25mm, bottom=25mm,
  left=18mm, right=18mm,
  headheight=0pt, headsep=0pt,
  footskip=10mm
]{geometry}

\usepackage{amsmath,amsfonts}

\usepackage{graphicx}
\usepackage{booktabs}
\usepackage{threeparttable}
\usepackage{caption}
\usepackage{xcolor}
\usepackage{gensymb}
\usepackage{colortbl}
\usepackage{textcomp}
\usepackage{csquotes}
\usepackage{placeins}
\usepackage{ragged2e}  

\usepackage{cite}
\usepackage{hyperref}
\hypersetup{
  colorlinks=false,
  pdfborder={0 0 0}
}

\usepackage{titlesec}
\titleformat{\section}
  {\normalfont\bfseries\fontsize{11}{13}\selectfont\centering}
  {\thesection.}{0.5em}{}
\titleformat{\subsection}
  {\normalfont\bfseries\fontsize{10}{12}\selectfont}
  {\thesubsection}{0.5em}{}
\titlespacing*{\section}   {0pt}{10pt}{6pt}
\titlespacing*{\subsection}{0pt}{8pt}{4pt}

\usepackage[bottom]{footmisc}
\renewcommand{\footnoterule}{%
  \kern-3pt\hrule width\columnwidth height 0.4pt\kern3pt}

\newlength{\abstractwidth}
\begin{document}
\pagestyle{empty}

\twocolumn[{%
  \begin{minipage}{\textwidth}

    \begin{center}
      {\fontsize{16}{18}\selectfont
      Toward Generalizable Graph Learning for 3D Engineering AI
      \par}

      \vspace{2pt}

      {\fontsize{12}{13}\selectfont
      Explainable Workflows for CAE Mode Shape Classification and CFD Field Prediction
      \par}

      \vspace{12pt}

      {\fontsize{11}{15}\selectfont\bfseries
        Tong Duy Son, Kohta Sugiura, Marc Brughmans, Andrey Hense, Zhihao Liu, \\
        Amirthalakshmi Veeraraghavan, Ajinkya Bhave, Jay Masters, Paolo di Carlo, Theo Geluk
      }
    \end{center}

    \vspace{5pt}

    \noindent\hfill
    \begin{minipage}{\abstractwidth}
      \setlength{\parindent}{4mm}
      \setlength{\parskip}{0pt}
      {\fontsize{9}{11}\selectfont\justifying
      \noindent\hspace{4mm}%
        Automotive engineering development increasingly relies on heterogeneous
        3D data, including finite element (FE) models, body-in-white (BiW)
        representations, CAD geometry, and CFD meshes.
        At the same time, engineering teams face growing pressure to shorten development cycles, 
        improve performance and accelerate innovation. 
        Although artificial intelligence (AI) is increasingly explored in this
        domain, many current methods remain task-specific, difficult to
        interpret, and hard to reuse across development stages.
        This paper presents a practical graph learning framework for 3D
        engineering AI, in which heterogeneous engineering assets are converted
        into physics-aware graph representations and processed by
        Graph Neural Networks (GNNs). The framework is designed to support both classification and prediction tasks. 
        The framework is validated on two automotive applications: CAE
        vibration mode shape classification and CFD aerodynamic field prediction.
        For CAE vibration mode classification, a region-aware BiW graph supports explainable 
        mode classification across vehicle and FE variants under label scarcity.
        For CFD aerodynamic field prediction, a physics-informed surrogate predicts pressure 
        and wall shear stress (WSS) across aerodynamic body shape variants, 
        while symmetry preserving downsampling retains accuracy with lower computational cost.
        The framework also outlines data generation guidance that can help engineers identify which additional simulations
        or labels are valuable to collect next.
        These results demonstrate a practical and reusable engineering AI
        workflow for more trustworthy CAE and CFD decision support.

      \vspace{5pt}
      }
    \end{minipage}%
    \hfill\null

    \vspace{20pt}
  \end{minipage}
}]

{\renewcommand{\thefootnote}{}\footnotetext{The authors are with Siemens Digital Industries Software, Leuven, Belgium. Contact Email: son.tong@siemens.com.  
This work is part of the ROBUSTIFAI project (grant agreement No.
101212818) funded by Horizon Europe – the Framework Programme for Research
and Innovation. The work also benefited from the Flanders Innovation \&
Entrepreneurship  VLAIO funded project SATISFY.AI.}}


\input{introduction}

\input{background}

\input{framework}

\input{UseCaseA}

\input{UseCaseB}

\input{conclusion}


\bibliographystyle{ieeetr}
\bibliography{main}

\end{document}

%% file: introduction.tex

\section{Introduction}


Automotive product development relies heavily on simulation-driven engineering, 
where Computer-Aided Engineering (CAE) and Computational Fluid Dynamics (CFD) are used 
to evaluate structural, vibro-acoustic, and aerodynamic performance before physical 
prototypes are finalized \cite{vanderauweraer2007structural}.
In structural development, finite element (FE) models are used to study body stiffness, modal behavior, and noise, vibration, and harshness (NVH) performance, while aerodynamic simulations are used to predict drag, pressure distribution, wall shear stress, and flow behavior around the vehicle
\cite{ewins2000modal,cfd_automotive,heft2012introduction}.
These analyses are central to engineering decisions on lightweighting, body structural refinement, 
aerodynamic efficiency, and overall product performance.


Vehicle development generates multiple digital artifacts across design, simulation, testing, and validation workflows.
These artifacts include 3D CAD geometry, body-in-white (BiW) and trimmed-body representations, 
finite element models, CFD surface meshes with field data, and test and validation measurements.
In practice, these assets are distributed across different tools, teams, sensor layouts, 
and configuration definitions.
As a result, engineering knowledge is often fragmented across domains, making it difficult to 
reuse historical data and build AI workflows that operate consistently across the development process.

\begin{figure*}[t]
  \centering
  \includegraphics[width=\textwidth]{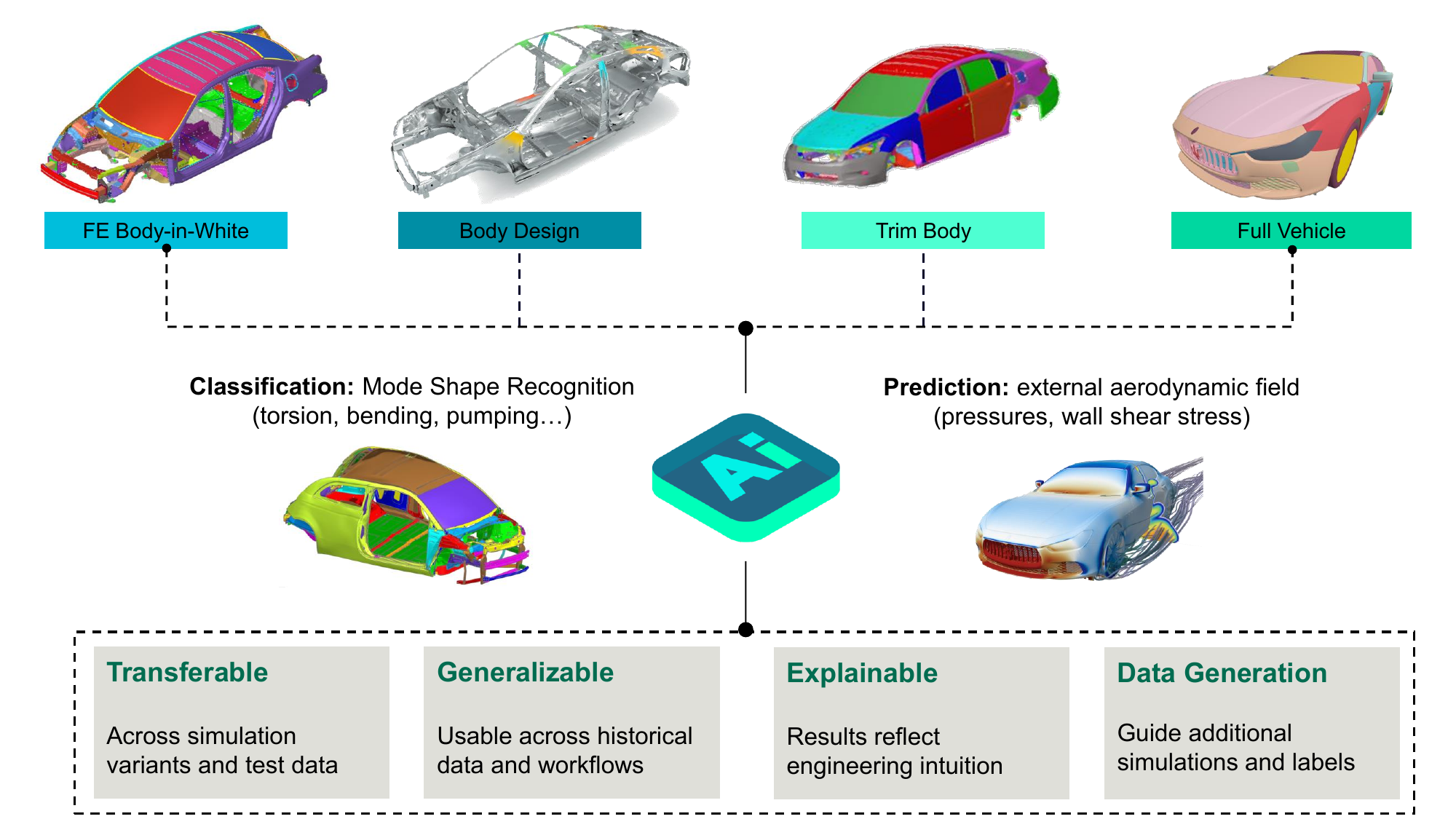}
  \caption{Overview of the proposed graph learning framework for 3D engineering AI. Heterogeneous 3D models are mapped into engineering-guided graph representations that support CAE mode shape classification, CFD aerodynamic field prediction, and data generation workflow.}
  \label{fig:main_diagram}
\end{figure*}


AI adoption is accelerating in the automotive industry because it offers a way to reduce repetitive expert effort and 
accelerate engineering decision-making.
In CAE structural dynamics, AI can support structural interpretation, mode assessment, cross-variant comparison, 
and linkage with operational behavior during the development process
\cite{tohmuang2025modegcn,Millan10112025}.
In particular, vibration mode shape classification (e.g., bending, torsion, pumping, and local modes) remains time consuming and often depends on specialist judgment, 
which limits scalability when many model variants or iterations must be evaluated.
In CFD, the motivation is complementary.
High fidelity aerodynamic simulation is expensive, and field prediction is often needed for design exploration, optimization, and trade-off analysis
\cite{bhatnagar2019prediction,drivaernet2024,drivaerstar2025}.
Recent graph-based surrogates have shown that mesh-aware learning can accelerate fluid-related prediction while preserving more geometric structure than purely grid-based methods
\cite{sanchez2020learning}.

Both CAE and CFD therefore share a common opportunity: they require learning from complex 3D spatial structures, yet they also require outputs that remain meaningful to engineers.

Existing 3D AI methods address parts of this problem, but each comes with tradeoffs.
Voxel and CNN approaches struggle with industrial geometric scale, point-based methods often weaken explicit topology, and conventional surrogates depend heavily on manual feature design
\cite{tompson2017accelerating,bhatnagar2019prediction,qi2017pointnet,forrester2008engineering}.
Physics-informed learning improves consistency with governing principles, and graph-based methods preserve relational structure more naturally than grids or unordered points, but most published workflows are still tailored to one dataset or one narrowly defined task
\cite{raissi2019physics,velickovic2018graph}.

For engineering deployment, four challenges remain especially important:
\begin{enumerate}
  \setlength{\itemsep}{0pt}
  \setlength{\parsep}{0pt}
  \item {Data requirement:} reliable models need to be built under limited data availability,  without requiring large numbers of simulations or costly large-scale testing campaigns.
  \item {Interpretability and explainability:} engineers need trustworthy models that not only predict or classify but also help connect data-driven inference with physical reasoning.
  \item {Reuse across variants:} models should remain useful across related vehicle programs, mesh layouts, and configuration changes rather than only within one fixed dataset.
  \item {Industrial practicality:} workflows must operate under heterogeneous data definitions, limited labels, and the need for engineering review rather than only benchmark accuracy.
\end{enumerate}


Despite rapid progress in industrial AI, current pipelines are still often node-specific, case-specific, or weakly connected to the engineering entities that engineers actually use in review.
They also provide limited guidance on where new simulations, labels, or variant studies are most valuable once a surrogate or classifier is deployed.
This gap is especially important in automotive development, where CAE, CFD models and test data are closely related in practice but remain difficult to reuse consistently across variant programs and workflow stages.


This paper addresses that gap by formulating a graph-based framework.
The key idea is to transform 3D engineering inputs into graph
representations whose nodes, edges, and attributes retain physically
meaningful structure while remaining flexible enough to support different
downstream tasks.
The proposed framework is physics-informed through engineering intuition
embedded in graph construction, feature definition, pooling design, and task
formulation.
It combines a shared engineering representation strategy with use case specific
graph neural network (GNN) architectures.
The framework is validated on two representative engineering use cases: BiW
mode shape classification and aerodynamic field prediction.
Although the present demonstrations focus on FE-derived BiW and CFD surface
data, the same formulation is intended to be extensible to other engineering
assets, such as CAD and trimmed-body representations, when they can be
mapped into physically meaningful graph regions and relations.

The main contributions of this work are as follows:
\begin{enumerate}
  \setlength{\itemsep}{0pt}
  \setlength{\parsep}{0pt}
  \item Graph-based engineering workflows for heterogeneous 3D assets, demonstrated on BiW data and CFD surface data, and extensible to CAD and trimmed body assets.
  \item Region-aware BiW graphs and a hierarchical graph-attention explainable mode shape classification across BiW and FE, testing variants under severe label scarcity.
  \item Physics-informed aerodynamic surrogates based on surface graphs for efficient pressure and WSS field prediction, enabled by symmetry-aware graph preprocessing.
  \item An integrated workflow perspective in which explainability and uncertainty analysis support data generation in engineering practice.
\end{enumerate}

Explainability and uncertainty are considered in this work from a workflow perspective.
In particular, data generation is framed as a means of prioritizing future simulations or labels, rather than as a fully validated
closed-loop contribution. Finally, although other AI architectures, particularly transformer-based models, may
also support similar objectives, the present work focuses on GNNs because they
are well suited to limited data settings and allow prior physical information
to be incorporated through graph construction and message passing.

The remainder of this paper is organized as follows.
Section~2 reviews related works on CAE mode shape classification, CFD surrogate modeling, and
3D graph learning.
Section~3 presents the proposed AI framework.
Sections~4 and~5 present the main results of CAE and CFD use cases, respectively.
Section~6 discusses and concludes the paper.

%% file: background.tex

\section{Background}

This section reviews the technical background and prior work most relevant to
3D AI applications in CAE and CFD.
It also considers the present study within broader discussions on reusable and
domain-adapted AI frameworks in computational science, where clarity of scope
and rigor of evidence remain especially important \cite{choi2025foundationcs}.

\subsection{CAE 3D Mode Shape Recognition}

3D mode shape recognition is a long standing desired capability in automotive CAE because
engineers must interpret finite element models, understand modal deformation
patterns, and relate them to NVH performance, stiffness targets, and design
trade-offs \cite{vanderauweraer2007structural,tao2008vehicle,ewins2000modal,Millan10112025}. 
In addition, there is an emerging need for mode shape recognition to extract modal 
KPIs from CAE modal analysis result files for NVH data-driven modelling.

Traditional workflows rely on simulation outputs followed by manual expert
classification of torsion, bending, pumping, and local structural modes, often
requiring repeated review across vehicle variants and development cycles.
Although valuable, this process is expensive when multiple body architectures,
mesh densities, and subsystem changes must be compared under tight program
schedules.

Early attempts to automate mode interpretation relied on rule-based systems,
modal parameter estimation, modal assurance criterion (MAC), and hand-crafted
features extracted from structural simulations \cite{ewins2000modal,gioia2020validation}.
Classical supervised learning methods and other feature-based classifiers
improved automation but still depended heavily on expert-defined preprocessing
and often remained tied to a particular vehicle or mesh representation.
More recently, graph-based learning has emerged as a promising direction for
structural data because it can preserve regional connectivity and modal
relationships more naturally than flat feature vectors or purely geometric
point sets \cite{velickovic2018graph}.
The recent automotive study showed that graph
convolutional networks can support structural mode classification directly from
engineering representations, highlighting the practical potential of graph
learning in this domain \cite{tohmuang2025modegcn}.

However, three limitations remain.
First, most current approaches have limited cross-model transfer, especially
between different body platforms and between simulation and testing data.
Second, explainability remains insufficient for engineering review, since high
classification performance alone does not reveal why a mode is assigned to a
particular structural category.
Third, existing approaches rarely provide a broader framework for exploiting
historical engineering data across vehicle programs, even though such reuse is
central to industrial value.

\subsection{CFD External Aerodynamic Surrogate}

3D aerodynamic prediction is another major target for automotive AI because CFD
remains essential for drag reduction, flow control, thermal management, and
energy-efficiency optimization \cite{cfd_automotive,heft2012introduction}.
In industrial practice, high fidelity CFD can require hours per simulation,
making rapid design exploration or optimization expensive when various body variants, ride
heights, or operating conditions must be evaluated.
This has motivated a broad class of surrogate models that aim to approximate
surface fields and integrated aerodynamic coefficients while preserving
sufficient fidelity for engineering use.

Classical surrogate approaches include reduced order models and response
surface methods, which can be efficient but often struggle when geometry, flow
regime, or output dimensionality becomes highly nonlinear
\cite{forrester2008engineering}.
With the growth of deep learning, CNN based approaches have been used for flow
field approximation, but they generally require regularized representations and
are therefore less natural for high fidelity irregular vehicle surfaces
\cite{tompson2017accelerating,bhatnagar2019prediction}.
Mesh and graph based approaches, including MeshGraphNet models, have
shown that message passing can better preserve local geometric structure and
field interactions on irregular engineering domains
\cite{sanchez2020learning}.
Physics-informed learning further strengthens this direction by embedding
consistency with governing principles directly into the learning process
\cite{raissi2019physics}.

Recent automotive benchmark efforts have clarified this landscape.
DrivAerNet++ established a large scale multimodal basis for aerodynamic
learning on realistic car geometries, while DrivAerStar presented a
high fidelity aerodynamic prediction framework built on industrial scale
meshing (12 million cells) and CFD solver outputs \cite{drivaernet2024,drivaerstar2025}.
CarBench further highlights the importance of benchmark-driven evaluation of
neural surrogates for 3D car aerodynamics, including graph-based,
transformer-based, and hybrid model families \cite{carbench2025}.

Despite this progress, three gaps remain especially relevant.
First, CFD data generation remains expensive, creating a need for efficient
strategies to determine where additional simulations will most improve a model.
Second, many approaches remain tightly coupled to a specific dataset,
benchmark, or configuration, which limits transfer to new geometries and
engineering contexts.
Third, interpretation remains difficult, particularly when engineers need to
understand not only predicted values but also the surface regions and flow
structures driving those predictions.

\subsection{3D Engineering Data Representation and Learning}

The design of AI for engineering data depends strongly on representation.
Voxel and grid based methods offer direct compatibility with standard CNN
operations, but they often sacrifice geometric efficiency and physical detail
when complex 3D engineering objects are mapped onto regular domains
\cite{tompson2017accelerating,bhatnagar2019prediction}.
Point based methods such as PointNet and PointNet++ avoid explicit meshing and capture 
local geometric structure through hierarchical grouping, yet they do not provide a 
mechanism to incorporate engineer defined physical relations such as FE element adjacency, 
component level coupling, or load path connectivity.
\cite{qi2017pointnet}.
Mesh and graph based methods are well suited to engineering assets because
they preserve adjacency, connectivity, symmetry, and multiscale interactions
through explicit node-edge structure
\cite{battaglia2018relational}.
Transformer based learning is also increasingly relevant for 3D engineering
tasks, particularly in settings where long range dependencies, global context,
and multi-region interactions are important \cite{carbench2025}.
More broadly, recent works suggest that graph neural network, transformer, and
hybrid architectures exhibit different accuracy-efficiency tradeoffs depending on 
mesh scale and prediction target.

Across these representation families, the most important criteria for
engineering use are structure preservation, physical meaning,
interpretability, scalability, transferability, and suitability for sparse or
expensive labels.
Accordingly, the present work focuses on graph-based learning not as an
exclusive solution class, but as a practically motivated and structurally
appropriate choice for the engineering settings considered in this research.

\subsection{Explainable and Trustworthy AI in Engineering}

Black-box AI remains difficult to adopt in CAE and CFD practice because
engineering decisions are not based on prediction accuracy alone.
Engineers must understand which structural regions, geometric features, or flow
patterns influence a prediction before they can trust the model in design
reviews or deploy it in iterative workflows.
This has motivated the use of explainability tools such as attention
visualization, saliency analysis, sensitivity analysis, and integrated gradients \cite{velickovic2018graph}.
Although such techniques are useful, they often remain primarily data-driven
and are not always connected back to domain knowledge, engineering
terminology, or physically meaningful regional decomposition.

Trustworthiness also requires uncertainty awareness.
In engineering settings, uncertainty quantification is not only useful for
measuring confidence; it also supports expert review and decisions about
further data generation when predictions are ambiguous or out of distribution.
Bayesian approximations and related uncertainty aware deep learning methods
provide one path toward this goal \cite{gal2016dropout}.
Human-in-the-loop review is therefore not a fallback mechanism but a core part
of responsible deployment, especially when model outputs are used to guide
simulation budgets, structural interpretation, or aerodynamic design changes.
For this reason, explainability, uncertainty awareness, and domain grounded
interpretation must be treated as central properties of engineering AI rather
than optional results.

%% file: framework.tex

\section{AI Framework}

\begin{figure*}[t]
  \centering
  \includegraphics[width=\textwidth]{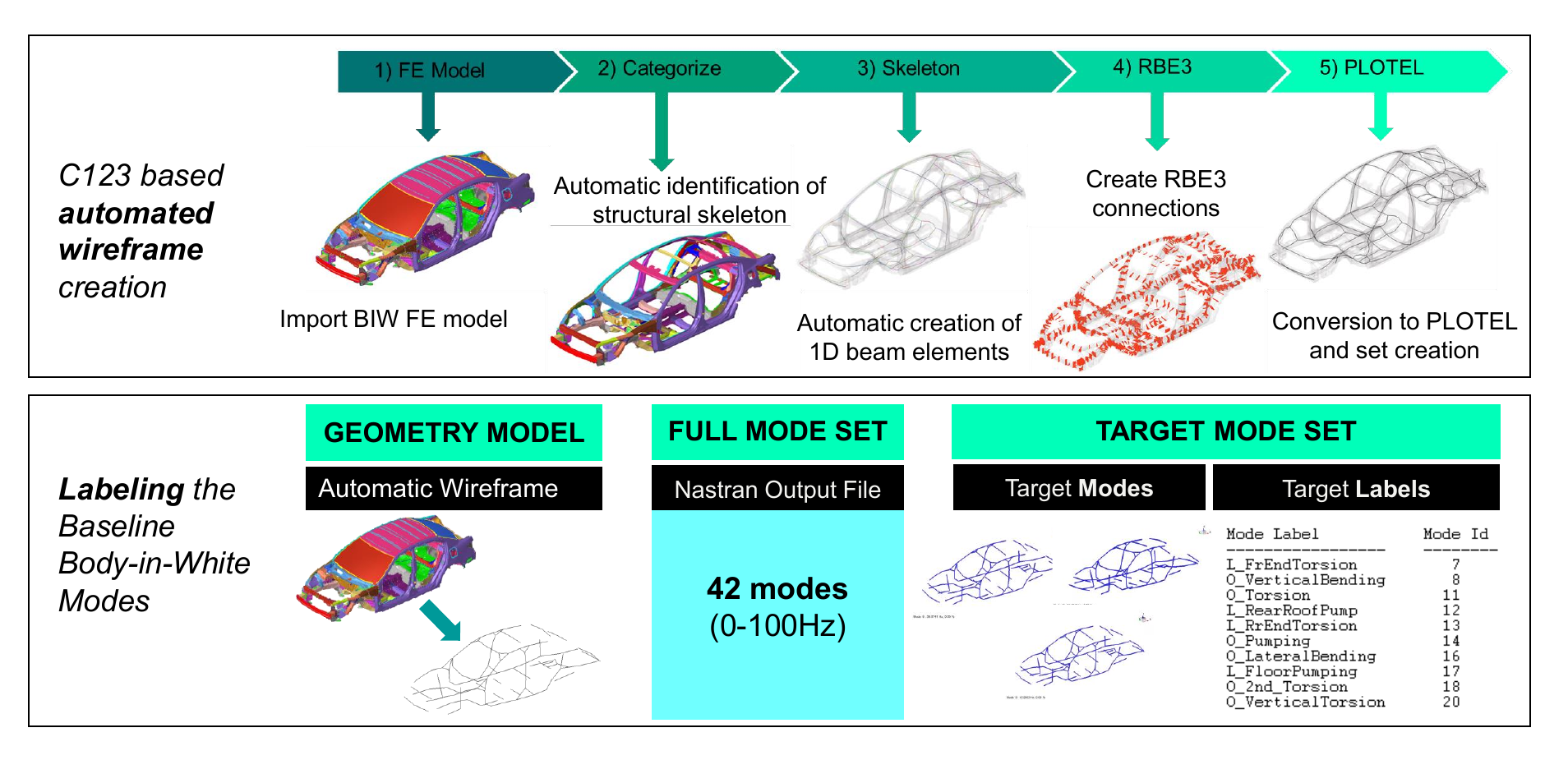}
  \caption{Overview of data generation and mode shape labelling pipeline.}
  \label{fig:uc1_data}
\end{figure*}

This section defines the common formulation used across the two use cases.
Rather than treating CAE mode classification and CFD field prediction as
completely unrelated AI problems, the proposed approach formulates both as
learning problems on physics-aware graphs derived from heterogeneous 3D
engineering data.
In this paper, the foundation-style aspect lies in the reuse of a common graph
abstraction and graph learning philosophy across engineering variants and task
types, rather than in claiming one universal pretrained model or one shared
cross-domain network architecture.

\subsection{Graph Formulation and Learning Pipeline}

Let an engineering sample be represented as a graph
$$
G = (\mathcal{V}, \mathcal{E}, \mathbf{X}, \mathbf{R}),
$$
where \( \mathcal{V} \) and \( \mathcal{E} \) denote nodes and edges,
\( \mathbf{X} \) contains node attributes, and \( \mathbf{R} \) contains edge
or relation attributes.
This formulation is intentionally generic so that it can represent FE models,
BiW regional skeletons, CAD-derived surfaces, and CFD meshes under one
abstraction.
Depending on the use case, nodes may represent structural regions or surface
samples, while edges encode physically meaningful relations such as adjacency,
load-path coupling, local neighborhood, or symmetry.

After graph construction, a use-case-specific graph neural network (GNN)
encoder is used to learn latent node embeddings through message passing.
Each node is updated from its own attributes together with information
aggregated from neighboring nodes and their relations, so the representation
captures both local descriptors and higher-level interactions.

At layer \(l\), a generic message passing update can be written as
$$
\mathbf{m}_{ij}^{(l)} = \phi^{(l)}\!\left(
\mathbf{h}_i^{(l)}, \mathbf{h}_j^{(l)}, \mathbf{r}_{ij}
\right),
$$
$$
\mathbf{h}_i^{(l+1)} = \psi^{(l)}\!\left(
\mathbf{h}_i^{(l)},
\square_{j \in \mathcal{N}(i)} \mathbf{m}_{ij}^{(l)}
\right),
$$
where $\phi^{(l)}$ is the message function and $\psi^{(l)}$ is the node update function, and
\( \square \) denotes an aggregation operator such as sum, mean, or
attention-weighted aggregation.
After \(L\) layers, the encoder produces node embeddings that are passed to one
or more task-specific output heads \cite{battaglia2018relational}.

For graph-level classification, node embeddings are pooled into a graph
representation
$$
\mathbf{z}_G = \mathrm{Pool}\left(\{\mathbf{h}_i^{(L)}\}_{i \in \mathcal{V}}\right),
$$
and decoded for class prediction.
If the task benefits from explicit analytical descriptors, this pooled graph
embedding can also be fused with engineered global features derived from the
same graph construction, as in the region-aware structural descriptors used in
Use Case A.
For node-level regression, each node embedding is decoded directly,
$$
\hat{\mathbf{y}}_i = f_{\mathrm{node}}(\mathbf{h}_i^{(L)}),
$$
which is suitable for distributed outputs such as pressure or wall shear
stress.
More generally, the framework also supports coupled multi-head settings in
which one graph sample produces several related outputs, such as hierarchical
classification labels and location descriptors.
Under this formulation, task variation is handled mainly through graph
construction, supervision, architecture choice, engineered feature fusion, and
output-head design.
The commonality across use cases is therefore at the level of representation
and engineering-AI principles, not at the level of a single cross-domain
network architecture.

\subsection{Engineering-Oriented Design Requirements}

From a practical engineering perspective, three requirements guide the
framework.
First, the graph must remain interpretable so that inputs and predictions can
be traced back to recognizable regions, components, or surface areas already
used in engineering review.
Second, the same overall formulation should support different task types,
including classification for structural interpretation and regression for
aerodynamic field prediction, without requiring a completely different
representation philosophy for each application.
Third, the workflow should remain useful under realistic industrial conditions,
where data are heterogeneous, labels may be limited, and engineers need
traceable predictions and explanation rather than only a numerical output.

GNNs are well suited to this role because they preserve irregular engineering
structure while allowing domain knowledge to be embedded through node
definition, edge relations, and task-specific supervision.
This engineering alignment is also what makes transfer learning more natural.
When the graph is defined in terms of recognizable physical regions rather than
raw mesh indices, the learned representation becomes less dependent on a single
simulation model and more reusable across related vehicle variants, mesh
layouts, and engineering configurations.
This is also important for engineering deployment.
When explanation is mapped back to engineering regions or parameters, the model
can indicate which structural zones or flow regions are driving the result and
therefore support human review more directly.
The same workflow logic can be extended one step further.
When region- or surface-level explanations are combined with uncertainty
estimates, the model can help prioritize which new simulations, labels, or
variant studies are most valuable to run next.
In this paper, that uncertainty-guided data-generation role is introduced as a
practical engineering workflow extension rather than a fully validated end-to-
end active-learning loop.
In Use Case A, this supports hierarchical BiW mode classification on a
canonical structural graph.
In Use Case B, it supports surface-field prediction on downsampled aerodynamic
meshes.
These two examples show how the same graph learning logic can be adapted to
practical CAE and CFD workflows without changing the overall framework.

%% file: UseCaseA.tex

\section{Use Case A: CAE Mode Shape Classification}

This use case addresses automatic classification of automotive body vibration
mode shapes.
Traditional approaches rely on handcrafted features or data-driven methods that
treat structural components as generic nodes.
Recent GNNs have shown promise by learning spatial
relationships through message passing, but they do not inherently encode
structural semantics.
Without additional engineering structure, the network must infer distinctions
such as floor, roof, pillar, or rail behavior purely from data, which becomes
difficult under severe label scarcity and makes the resulting predictions harder
to interpret in engineering review.

The proposed region-aware architecture addresses these limitations by combining
learned graph representations with analytical descriptors derived from
3D body-region groupings.
While standard global pooling aggregates node features uniformly, the present
approach performs semantic-aware aggregation aligned with the hierarchical
organization of the BiW.
This design is motivated by how engineers interpret mode shapes in practice: they do not focus on individual nodes, but rather on the response of key structural regions and their couplings.
For example, bending modes tend to concentrate response in rails and pillars, while pumping
behavior localizes more strongly in floor or roof structures.
Compared with loss-driven physics-informed formulations, the engineering
knowledge here is embedded directly into the architecture through
differentiable analytical operations rather than through additional physical
constraint terms.

The application focus spans both early-stage BiW development models and more detailed FE variants, 
where engineers repeatedly inspect modal analysis response to identify torsional, bending, pumping, 
and local deformation patterns.
Under this formulation, the main goal is not universal transfer, but robust
reuse across related vehicle programs that share a canonical structural
decomposition. The same region-based representation also creates a practical path toward
testing workflows, in which measured responses can be aggregated onto
the same skeletal topology.

\subsection{Problem Formulation}

\begin{figure}[tb]
  \centering
  \includegraphics[width=\columnwidth]{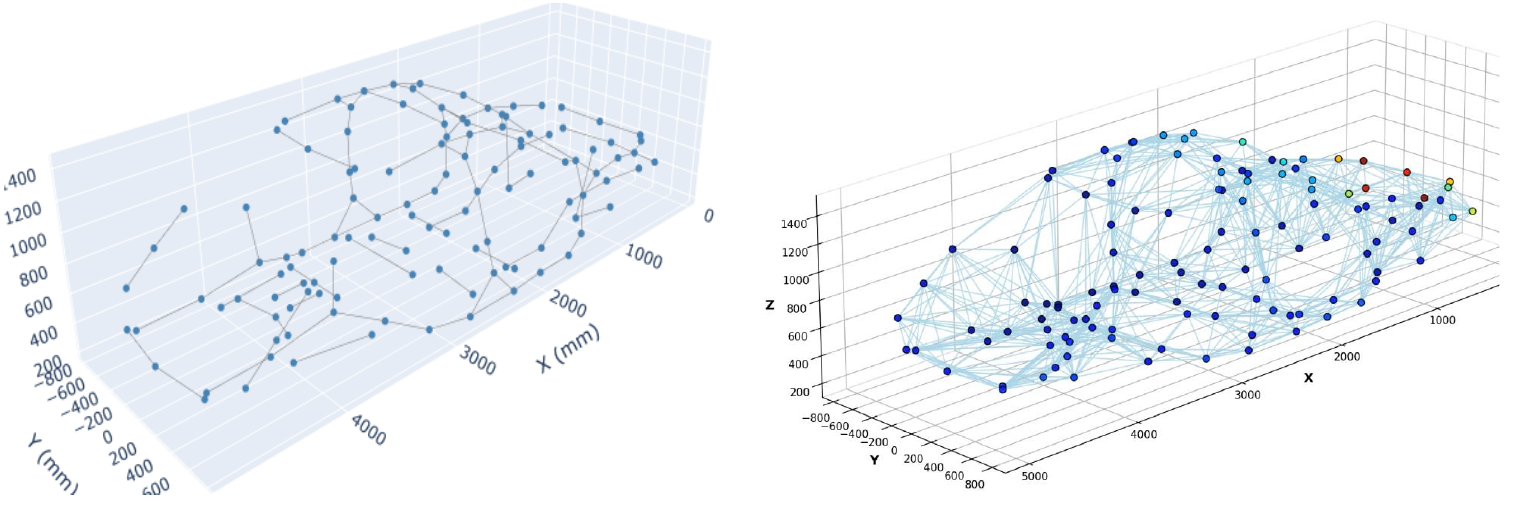}
  \caption{Graph construction from wireframe.}
  \label{fig:uc1_graph}
\end{figure}
\begin{figure}[tb]
  \centering
  \includegraphics[width=\columnwidth]{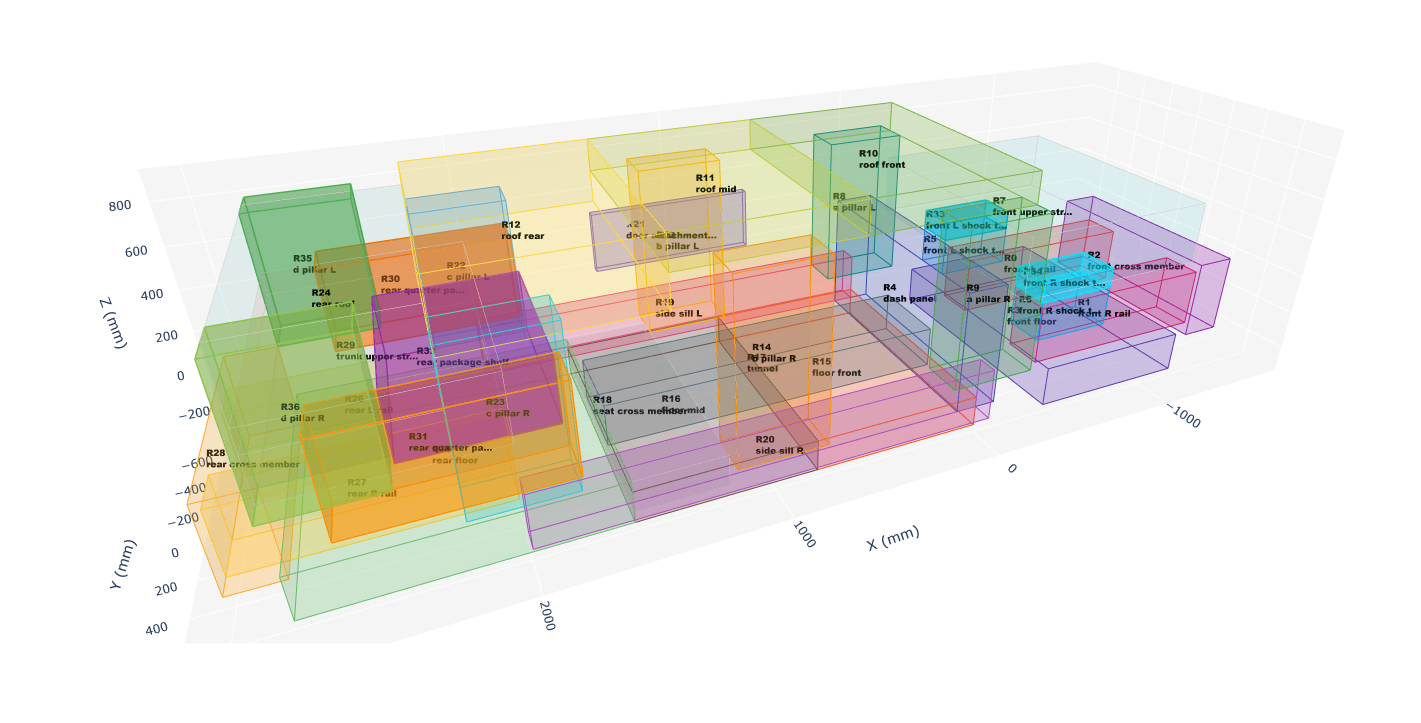}
  \caption{Canonical BiW regional decomposition used for region aware pooling and
  aggregation.}
  \label{fig:uc1_regions}
\end{figure}

Each mode shape is treated as one graph sample.
Starting from FE nodal displacement fields, the original model is aggregated
onto a canonical BiW skeleton composed of expert defined body engineering regions.
These regions correspond to consistent vehicle body structural regions such as front
rails, pillars, roof zones, floor zones, side sills, and rear members.
By mapping different FE models into the same regional decomposition, the
learning problem is recast from vehicle specific classification into
classification on a shared engineering graph.
This is also what makes the architecture relevant across simplified BiW
representations and more detailed FE layouts: the graph is tied to persistent
structural regions rather than to a fixed node numbering or discretization.

For this task, nodes represent BiW regions and edges encode four types of
engineering relations: structural adjacency, left-right symmetry,
longitudinal load path coupling, and vertical roof-floor coupling.
Each node carries aggregated regional features derived from the displacement
field, including mean and RMS displacement magnitude, directional response,
signed vertical behavior, and related quantities.
Each edge carries compact relational features that reflect edge type, regional
energy ratio, and phase agreement.

The prediction target follows the hierarchy used in NVH engineering review.
The model predicts a coarse Level-1 family, a fine-grained Level-2 subtype, and indicates potential
hybrid modes through confidence scores.
This structure reflects how engineers assess mode shapes in practice: first by
identifying the dominant global mechanism, then by refining the interpretation
into subtype and affected local regions.

\subsection{Dataset Generation and Labelling}

\begin{figure}[tb]
  \centering
  \includegraphics[width=\columnwidth]{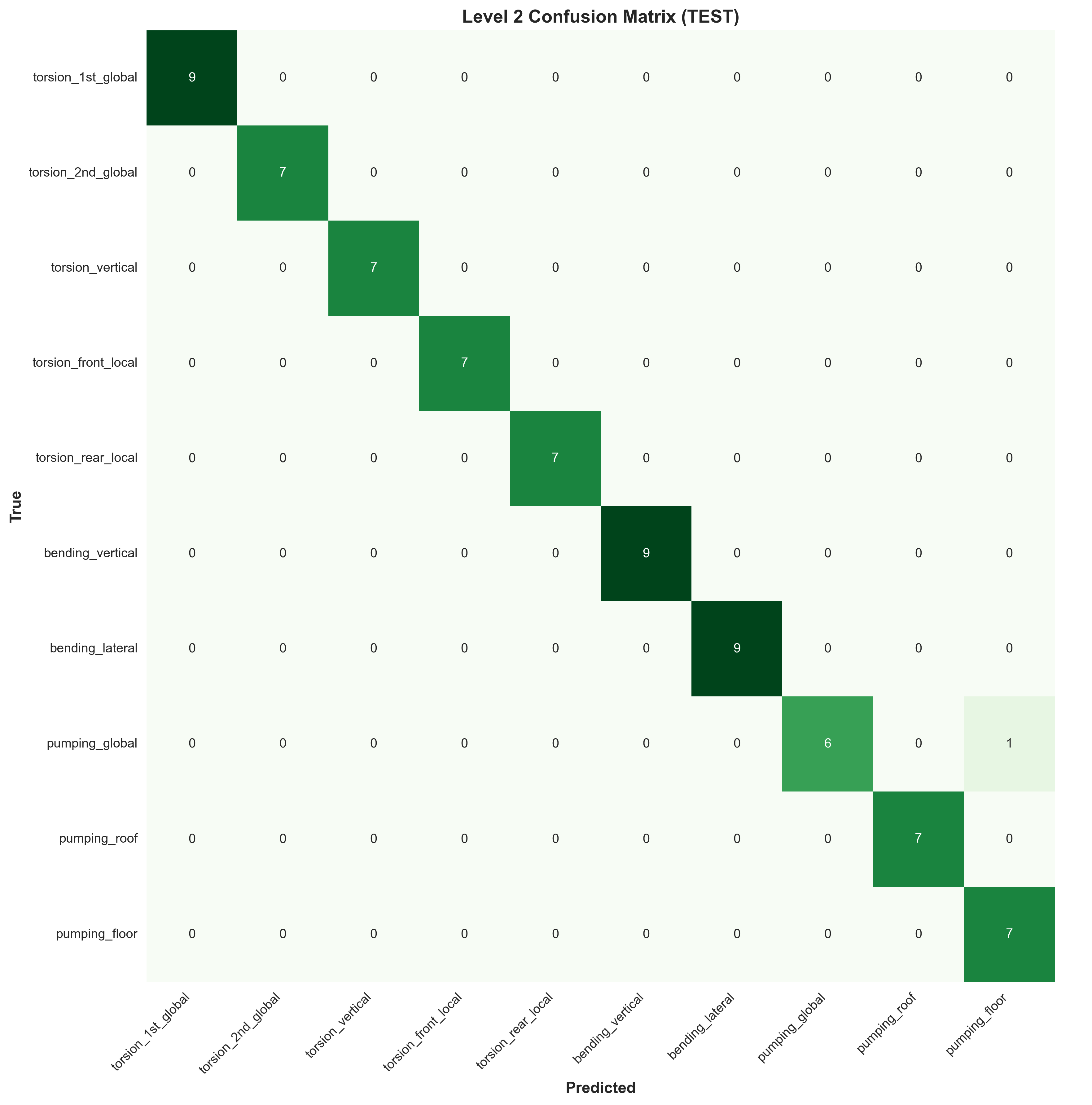}
  \caption{Confusion matrix for fine-grained Level-2 mode classification on the held-out multi-vehicle test set.}
  \label{fig:uc1_test_results}
\end{figure}

The dataset comprises computed BiW modes from four vehicle programs, of
which 326 were labeled by experienced NVH engineers and used for supervised
training and evaluation.
The vehicle set spans a mid-size sedan, compact sedan, luxury sedan, and sport
sedan.
Most labels come from the primary reference vehicle, while the additional
vehicles contribute only 9, 2, and 5 labeled modes, respectively.
This imbalance reflects the industrial scenario that motivated the work:
historical programs are well understood, whereas new programs begin with only a
small number of reviewed examples.

\begin{table}[tb]
  \centering
  \caption{Held-out test performance under different training strategies.}
  \label{tab:uc1_test_compare}
  \scriptsize
  \setlength{\tabcolsep}{3pt}
  \begin{tabular}{lccc}
    \toprule
    Training strategy & L1 Acc. (\%) & L2 Acc. (\%) & Comb. (\%) \\
    \midrule
    Reference-vehicle-only training & 90.8 & 81.6 & 85.3 \\
    Multi-vehicle training          & \textbf{100.0} & \textbf{98.7} & \textbf{99.2} \\
    \bottomrule
  \end{tabular}
\end{table}

To increase the amount of training data, we created variants of the original FE models with different vehicle characteristics, 
changing BiW beam stiffnesses, and materials. Figure~\ref{fig:uc1_data} illustrates the automatic data generation 
pipeline and wireframe creation from FE and BiW models using Siemens C123 technologies.
Because all variants share a common wireframe skeleton, 
MAC-based mode tracking between design variations is straightforward and reliable, 
providing a consistent basis for hierarchical label assignment across the expanded dataset. 
This consistency does not extend to cross-vehicle or cross-geometry transfer.
Expert labeling provides hierarchical supervision,
and physics-aware augmentation expands the limited labeled dataset before graph attention network training.
The relevant frequency band is approximately 0--100~Hz.
The evaluation follows a stratified 70/15/15 train/validation/test split by
vehicle and Level-2 class.

\subsection{Graph Construction and Learning Results}

\begin{figure}[tb]
  \centering
  \includegraphics[width=\columnwidth]{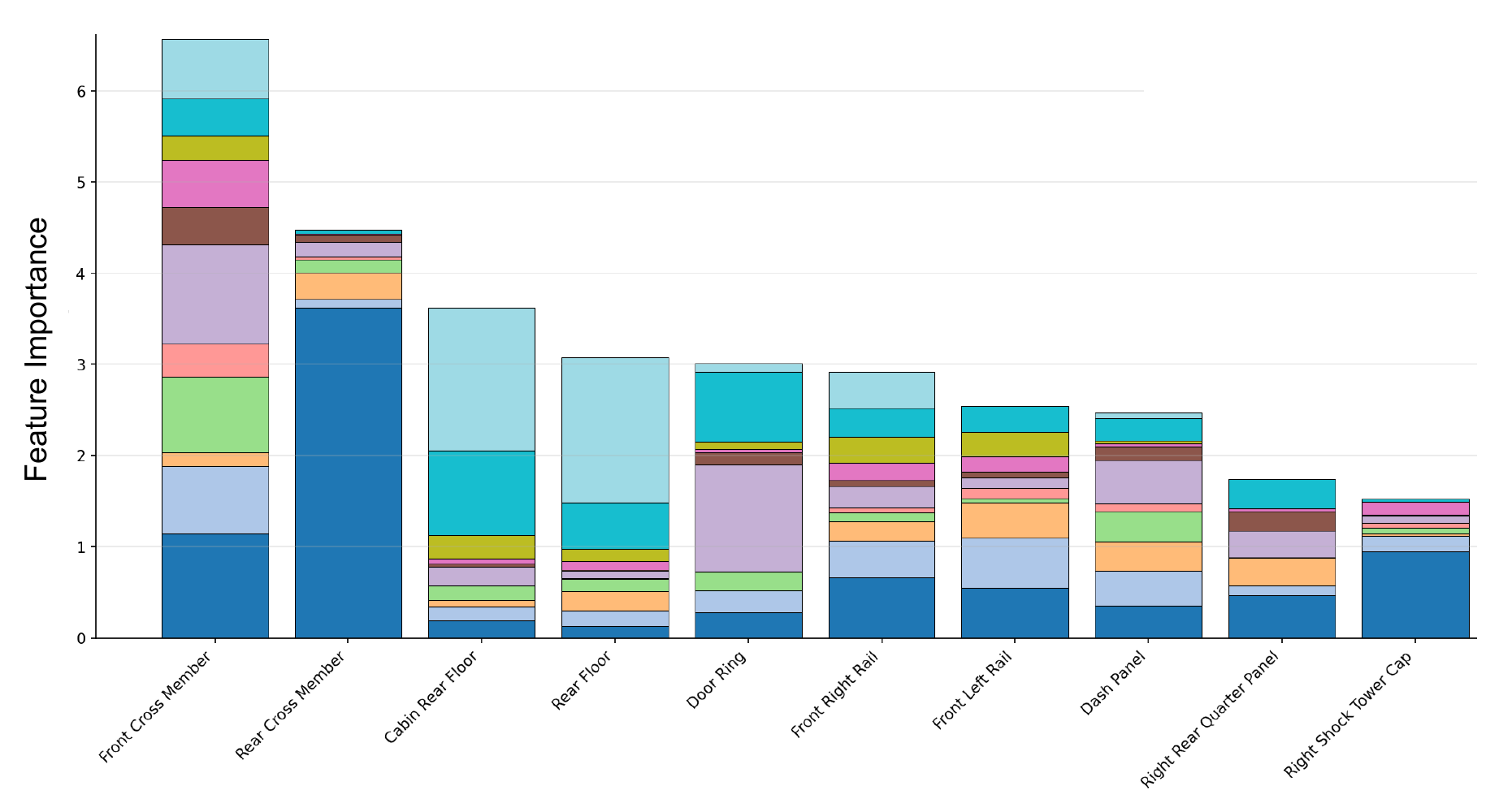}
  \includegraphics[width=\columnwidth]{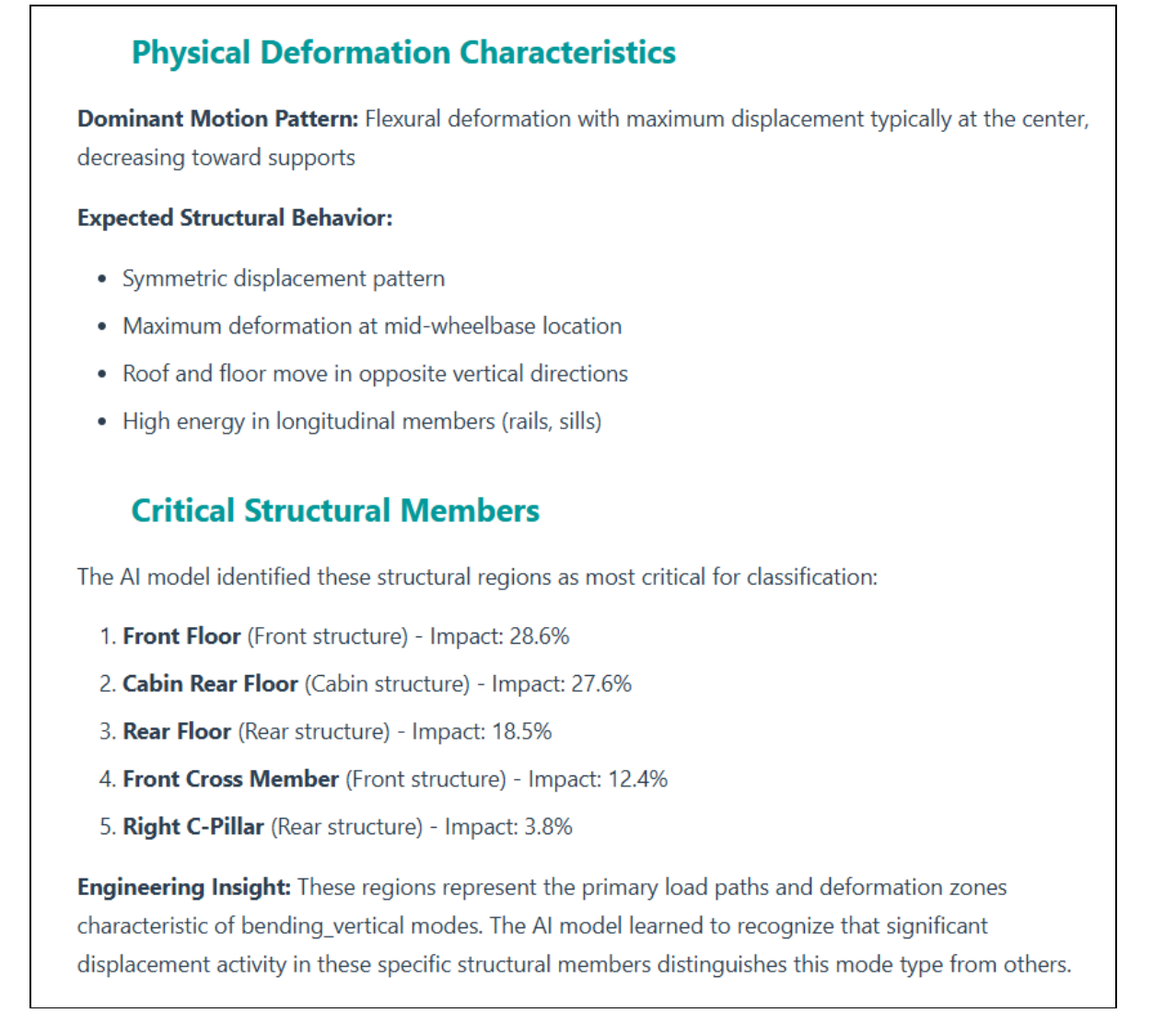}
  \caption{Illustrative explainability result for a mode shape classification, showing how
  model attribution can be mapped back to physically meaningful BiW regions.}
  \label{fig:uc1_explain}
\end{figure}

The graph topology is defined using engineering domain knowledge rather than
raw FE connectivity alone.
Fig.~\ref{fig:uc1_graph} shows the wireframe with the corresponding graph structure,
and Fig.~\ref{fig:uc1_regions} visualizes the canonical regional decomposition.
The key idea is to preserve physically meaningful coupling while removing
unnecessary dependence on the original mesh.

The classifier uses a 4-layer graph attention encoder with 8 attention heads.
To strengthen fine grained discrimination, the graph embedding is supplemented
with a region-aware pooling module that computes analytical scalars from the
regional responses, such as floor and roof energy fractions and global vertical energy
uniformity.
These features are especially useful for separating physically similar but
distinct classes, such as lateral bending versus floor pumping or roof pumping
versus floor pumping.
The network is trained with a multi-task objective that combines weighted
cross-entropy for Level-1 prediction with focal loss for Level-2 prediction.

The reported results are promising, especially considering the study scope:
the training dataset comprises 226 labeled modes across four vehicles, with severe label
imbalance and limited target-vehicle sample counts. A model trained only on the
reference vehicle already achieves strong in-domain performance, reaching 100.0\% at Level~2 on the held-out subset from that same
vehicle. However, its performance degrades when applied to the
other vehicle variants. When only 9, 2, and 5 labeled modes from the other three vehicles are added to the training dataset, the multi-vehicle training dataset model achieves 100\%
accuracy at Level~1, 98.7\% at Level~2, and 99.2\% on the combined metric on
the held-out multi-vehicle test set. Only one of 76 test samples is
misclassified, and that error remains within the
same broader pumping family. Validation performance is 98.2\%, and the hierarchical
consistency rate is 100\%, indicating that the joint heads preserve the intended
label structure.

In addition, the results show that the gains do not come from the GNN architecture
alone.
Multi-vehicle training with few target samples provides the significant improvement, followed by
cross-vehicle feature alignment, data augmentation, and
region-aware pooling.
Compared with single vehicle training, the full multi vehicle setting improves
Level-2 accuracy by 17.1 percentage points, showing that transfer performance
depends strongly on representation design and training protocol.

\subsection{Explainability and Interpretation}

\begin{figure}[tb]
  \centering
  \includegraphics[width=\columnwidth]{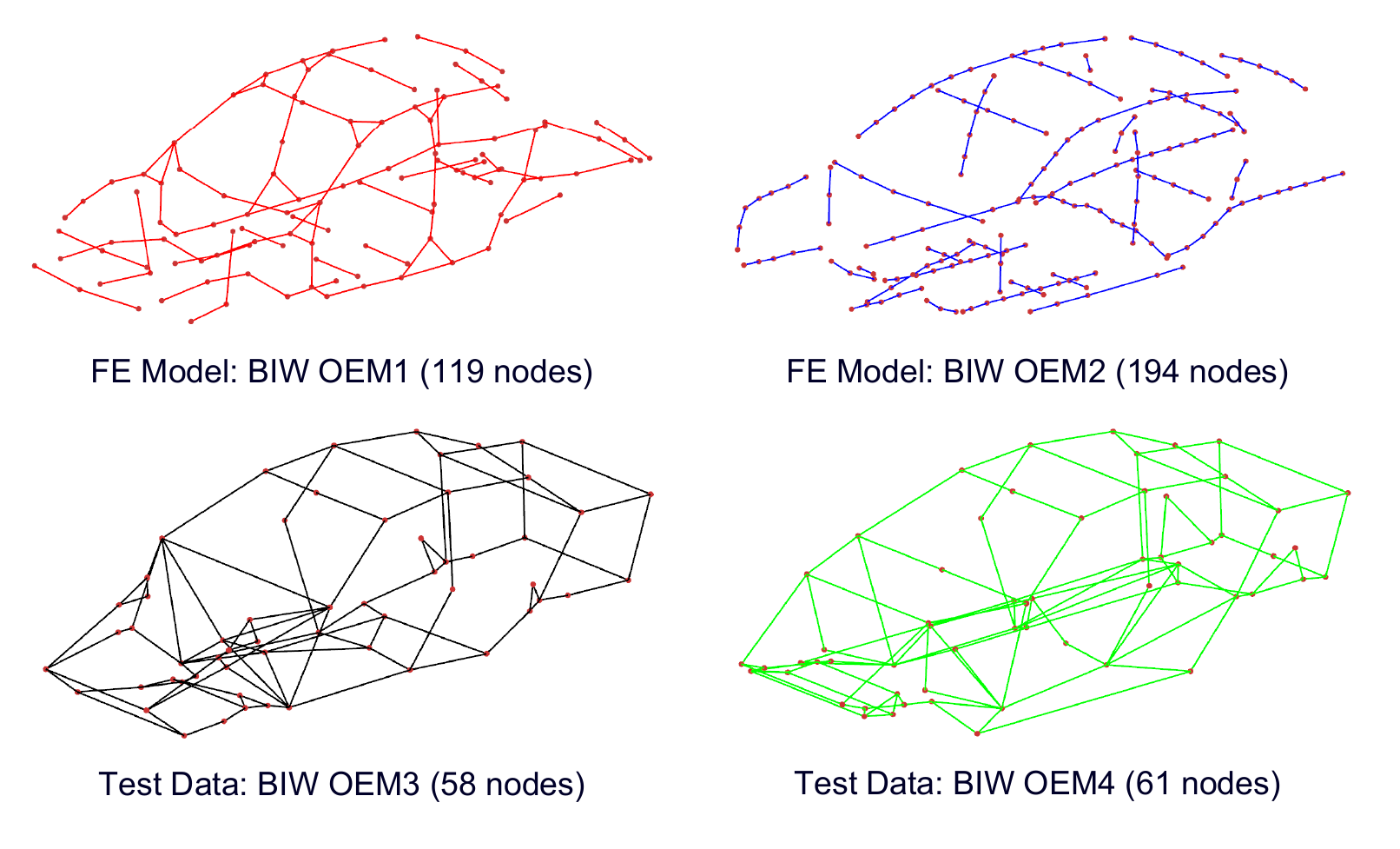}
  \caption{The four BiW structures with different body and node layout variants.}
  \label{fig:uc1_4models}
\end{figure}

Engineering deployment requires more than a correct label.
Vehicle dynamics and body engineers also need to understand which BiW regions drive a prediction and
whether those regions correspond to recognizable structural behavior.
Because the model operates on a small canonical regional graph, attribution can
be mapped directly back to engineering regions rather than to arbitrary FE node
indices.

Figure~\ref{fig:uc1_explain} illustrates this interpretation space.
Torsion modes emphasize pillar-to-sill couplings and diagonal structural paths,
bending modes focus more strongly on vertically coupled rails and pillars, and
pumping modes localize to floor or roof regions.
This kind of explanation is valuable in engineering review because it helps
analysts judge whether the model is using the same structural cues that would
be considered in manual classification.

\subsection{Generalizability and Transferability}

The most important outcome of this use case is the ability to achieve cross-vehicle transfer under severe label scarcity.
Only four vehicles and a total of 26 labelled mode shapes were available in the beginning.
To support reliable AI model training, one vehicle was selected for variant data generation using both a simulation-driven strategy guided by automotive engineering intuition and a complementary data-driven strategy.
Because all vehicles are represented through the same region-aware graph, the model operates in a consistent engineering space even when the underlying FE models or test-derived representations differ substantially.
This property is important in practice.
A classifier trained on a single vehicle may perform well within its own domain, yet remain inapplicable to another vehicle because of different input nodes or generalize poorly because of differences in overall 3D body structure.
By contrast, the proposed representation preserves regional semantics across vehicles and therefore supports transfer at the engineering level rather than only at the node or mesh level.

Figure~\ref{fig:uc1_4models} shows the four BiW structures used in the study.
In the reported experiments, the multi-vehicle aligned model correctly
classifies all evaluated samples in the small test
subsets of the other three vehicles at both coarse and fine-grained levels, whereas a single-vehicle
baseline degrades on those same targets.
This suggests that the canonical regional graph is doing more than reducing
input size: it is capturing structural invariants that remain meaningful across
vehicle programs.

The engineering value lies not only in the reported classification accuracy,
but also in the reusable region-aware representation, which remains meaningful
across discretization changes, layout differences, and related BiW variants.
Importantly, the last two target cases are based on test data and contain
substantially fewer sensor nodes than the FE-based cases, yet the model still
produces strong results when these data are mapped to the same regional
skeleton.
These results provide encouraging evidence that the proposed representation can
support transfer between simulation and test domains, although a broader and
more systematic simulation-to-test validation remains outside the scope of the
present study.
The findings also indicate that few-shot transfer is feasible across related
BiW architectures and FE variants with different discretizations and layouts,
despite the very limited number of target vehicle samples.
From a practical standpoint, this suggests that a new vehicle program could be
initialized with a small set of reviewed representative modes rather than a
large vehicle-specific labeled dataset.
These conclusions should nevertheless be interpreted in the context of the
dataset size, and broader coverage across all relevant structural regions would
be desirable in future work.

%% file: UseCaseB.tex

\section{Use Case B: CFD Aerodynamic Field Prediction}

\begin{figure}[tb]
  \centering
  \includegraphics[width=\columnwidth]{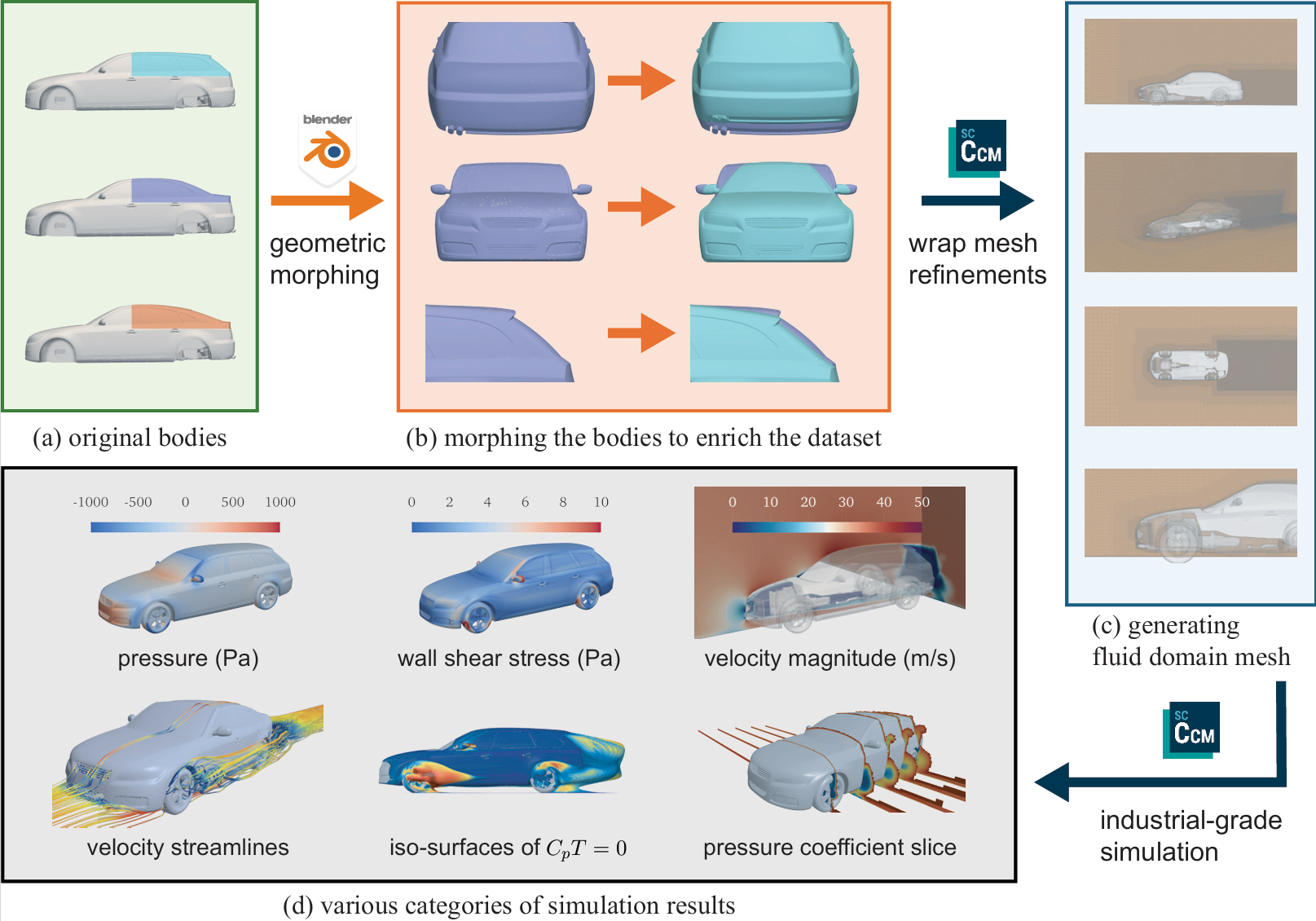}
  \caption{CFD data representation and preprocessing workflow, based on the DrivAerStar dataset generation \cite{drivaerstar2025}.}
  \label{fig:uc2_data}
\end{figure}

This use case evaluates the framework on external aerodynamic field prediction
from CFD-generated vehicle surface data.
While use case A focuses on graph-level structural classification, this use
case addresses dense node-level regression of aerodynamic surface fields.
High fidelity aerodynamic simulation remains one of the main computational
bottlenecks in vehicle concept development, particularly when many design
variants must be screened within short development cycles.

\subsection{Problem Formulation}

Each CFD sample is represented as a graph constructed from the vehicle surface
mesh.
The underlying data consist of high fidelity steady external aerodynamics
simulations in which each sample includes surface geometry together with local
pressure, wall shear stress (WSS), area, and surface normal information.
Because the original meshes contain approximately 375k surface vertices, direct
graph learning at full resolution is impractical for iterative training and
model selection.
The workflow therefore begins with symmetry-preserving downsampling to a graph
of approximately 15k nodes while retaining the bilateral structure that is
crucial for automotive exterior flow.

For the downsampled surface graph, each node carries geometry-aware features
such as normalized position, local area, surface normal, curvature, and
distance to the vehicle centroid.
Edges are defined through a local neighborhood graph with \(k=16\), enabling
message passing on the irregular surface topology while preserving local
geometric relations.
Within this representation, the model predicts surface distributed aerodynamic
quantities, particularly pressure and WSS.

\subsection{Dataset and Physics-Informed Learning}

\begin{figure}[tb]
  \centering
  \includegraphics[width=\columnwidth]{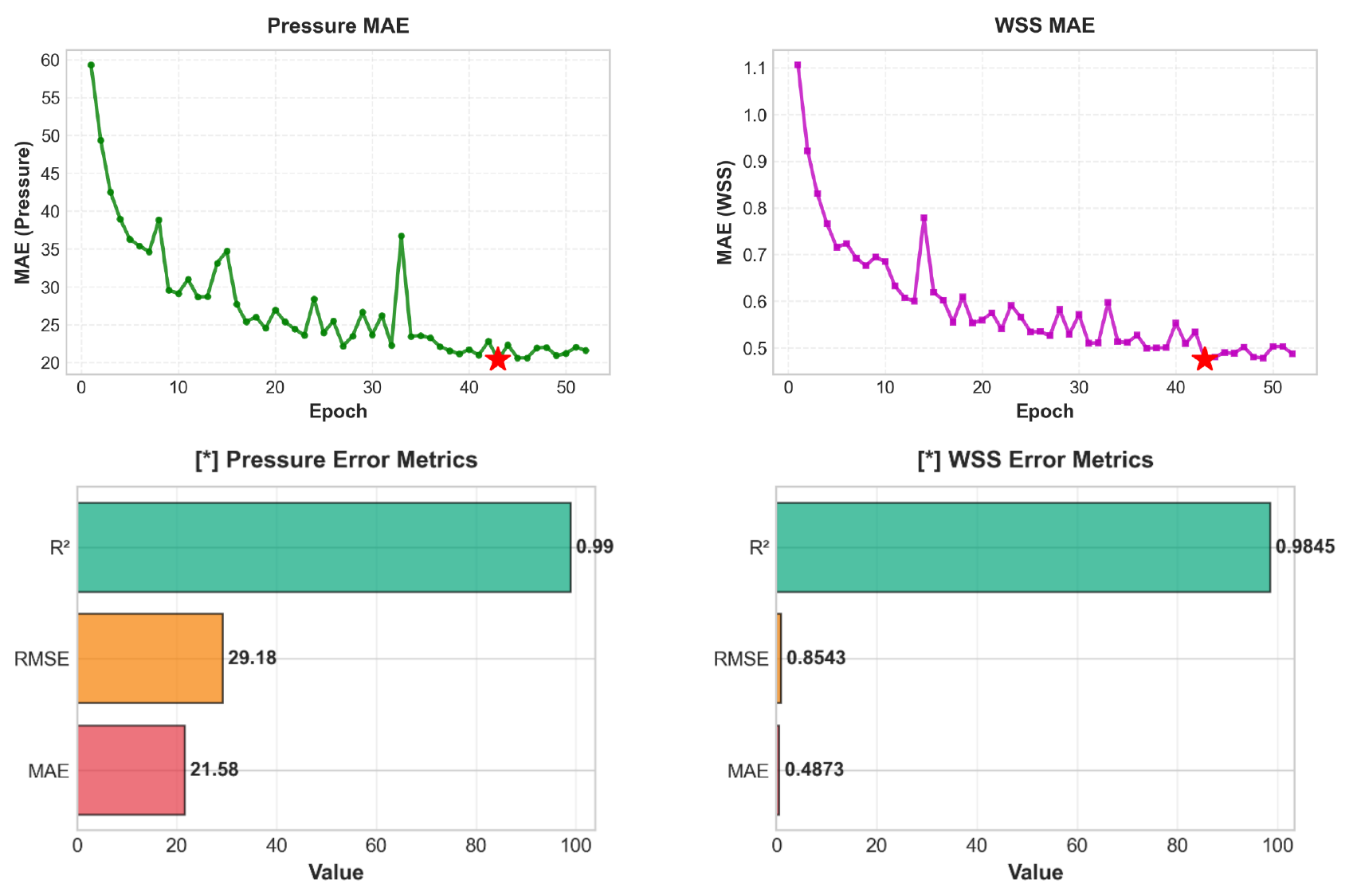}
  \caption{Training and validation performance of the physics-informed
  aerodynamic surrogate, showing training process and
  prediction accuracy for pressure and wall shear stress.}
  \label{fig:uc2_training}
\end{figure}

The aerodynamic use case is based on a dataset of approximately 10,000 CFD
simulations generated in STAR-CCM+ for three DrivAer-style body
configurations: Estateback, Fastback, and Notchback \cite{drivaernet2024,drivaerstar2025}.
The data are divided into 70\% training, 15\% validation, and 15\% test sets,
stratified by configuration, resulting in a held-out test set of about 1500
samples.
Compared with the CAE use case, this provides a much larger and more uniform
basis for assessing regression performance.

The predictive backbone is an attention-based AeroGraphNet with separate encoders 
for node and edge features, 6 message passing layers, 4 attention heads, and a decoder that outputs one
pressure value and a 3-component WSS vector per node.
The model size is approximately 2.3M parameters.
Training is designed to stabilize field prediction through a combination of
data-driven supervision and physics-informed regularization.
The loss design is one of the technical features.
Rather than relying solely on pointwise data fitting, the training objective
combines data fidelity with physically motivated regularizers, including
Bernoulli-style consistency, a mass conservation term, and WSS tangency to the
surface.
For the purposes of this paper, the key point is that these physics-informed
terms systematically improve predictive quality relative to the corresponding
data-only surrogate, albeit at the cost of increased training time.

\subsection{Training and Prediction Performance}

\begin{figure}[tb]
  \centering
  \includegraphics[width=\columnwidth]{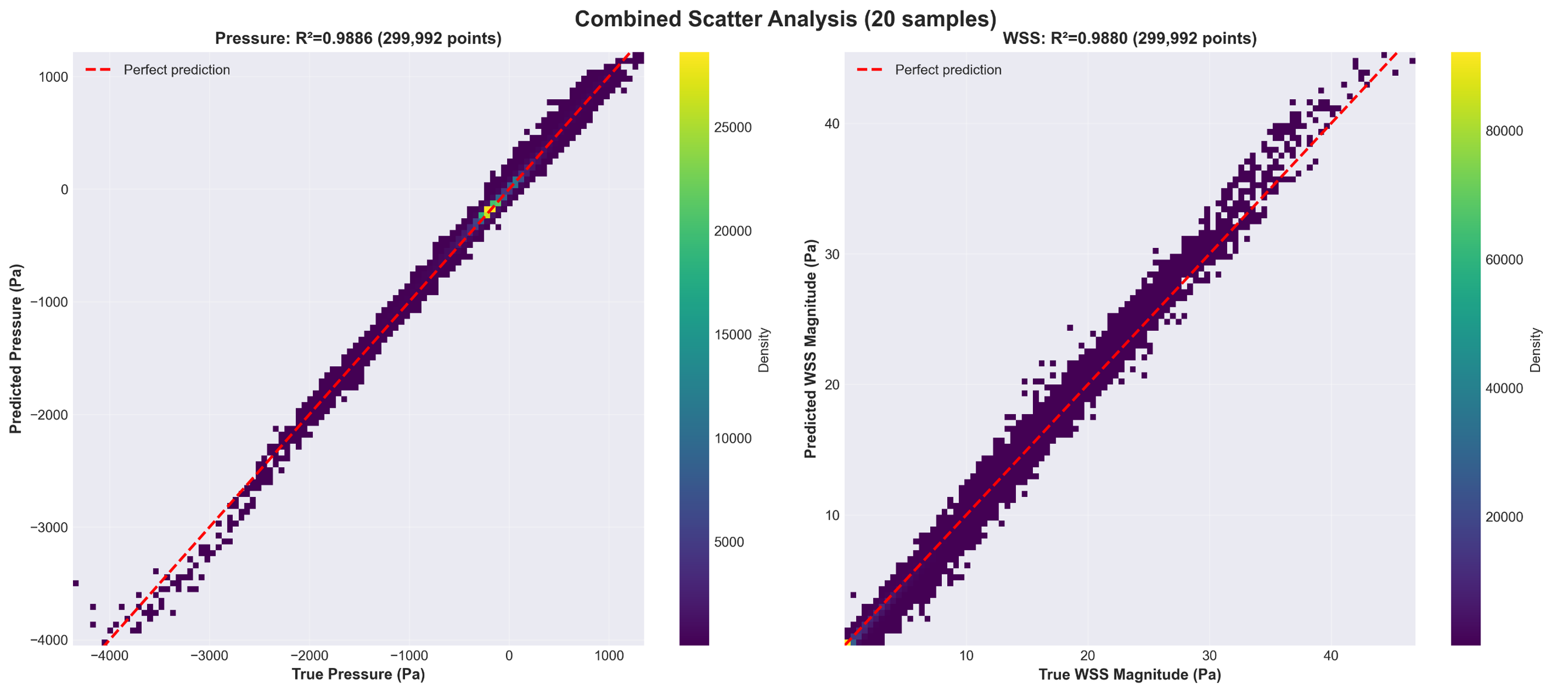}
  \caption{Quantitative performance comparison for aerodynamic field
  prediction: \(R^2 = 0.989\) for pressure and \(R^2 = 0.985\) for WSS,
  outperforming the evaluated baseline models.}
  \label{fig:uc2_plot}
\end{figure}

The results indicate strong field prediction performance.
The coefficient of determination \(R^2\) quantifies the proportion of variance in the reference data 
explained by the model, with values closer to 1 indicating stronger predictive accuracy.
The full physics-informed model achieves \(R^2 = 0.989\) for pressure and \(R^2 = 0.985\) 
for wall shear stress (WSS), together with a pressure mean absolute error (MAE) of 21.58~Pa and a WSS MAE of 0.49~Pa.
These results compare favorably with the evaluated GNN-based baselines, including MLP, GCN, MeshGraphNet, 
and a data-only version of the proposed architecture.
Relative to the data-only variant, the physics-informed formulation improves \(R^2\) by 6.6 points 
for pressure and 10.6 points for WSS.

The ablation studies show that symmetric downsampling is a central technical
enabler of the workflow rather than a simple preprocessing convenience.
Under the same 15k-node budget, the symmetry-preserving graph retains 99.8\% bilateral correspondence.
By comparison, random sampling reduces performance to 0.871 and 0.823, while curvature-based sampling 
reaches 0.912 and 0.891 for pressure and WSS, respectively.
Relative to the full 375k-vertex surface, the symmetric 15k-node graph preserves near comparable predictive
accuracy while substantially reducing training cost.

From an engineering perspective, computational speed is equally important.
Inference is reported at approximately 57~ms per sample on a GPU, representing a substantial 
speedup over the evaluated baselines.
This makes the model valuable not only as a high-accuracy surrogate, but also as a practical 
source of field-level feedback during rapid design iteration.

Additional studies indicate that edge-aware message passing is important, that \(k = 16\) provides the 
best trade-off between accuracy and computational cost, and that six message-passing layers outperform 
shallower alternatives.
Across the Estateback, Fastback, and Notchback subsets, pressure \(R^2\) remains consistently high, 
suggesting that the model generalizes across related body configurations rather than specializing to a 
single geometry type.

\begin{figure}[t]
  \centering
  \includegraphics[width=\columnwidth]{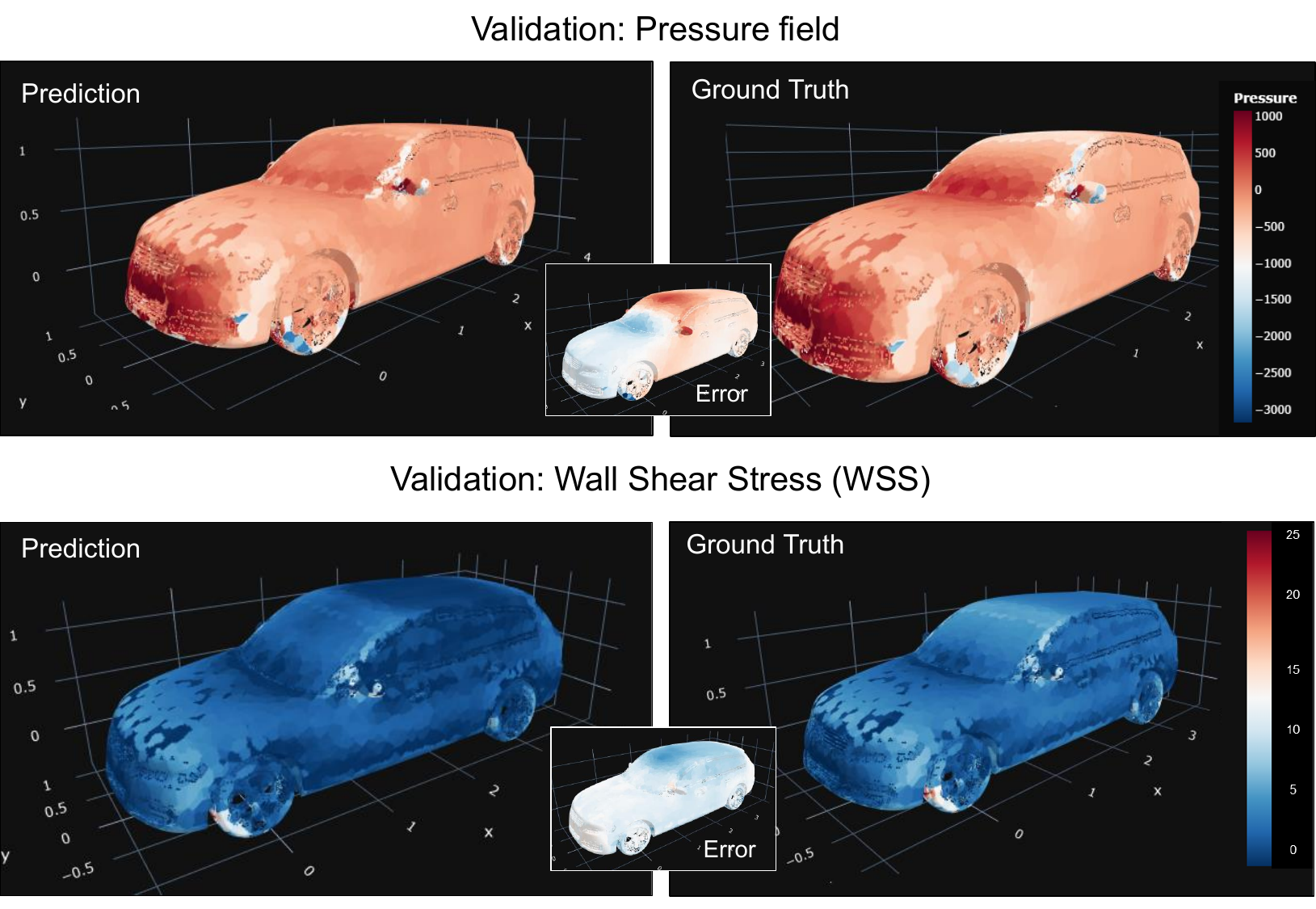}
  \caption{External aerodynamic field prediction results, comparing predicted surface fields with CFD reference data.}
  \label{fig:uc2_fields}
\end{figure}

The qualitative field visualizations are consistent with these quantitative results.
Front stagnation regions, rear pressure-recovery zones, and high-shear regions are reproduced with 
good fidelity, supporting the use of the model as a credible engineering surrogate for aerodynamic field prediction.

\subsection{Explainability and Workflow Development}

The aerodynamic model also provides an interpretable prediction space.
Attention based analysis and attribution maps indicate that the learned model
focuses on physically meaningful regions such as front stagnation areas,
rear end separation zones, and underbody flow sensitive regions.
This matters because explainability in engineering AI is only useful when it
can be translated back into recognizable flow structures and body zones that
engineers already use in design review.

This interpretability also makes the surrogate more actionable for geometry
refinement and design trade-off studies, as dominant pressure and WSS
regions can be linked back to recognizable body zones.

\begin{figure}[tb]
  \centering
  \includegraphics[width=\columnwidth]{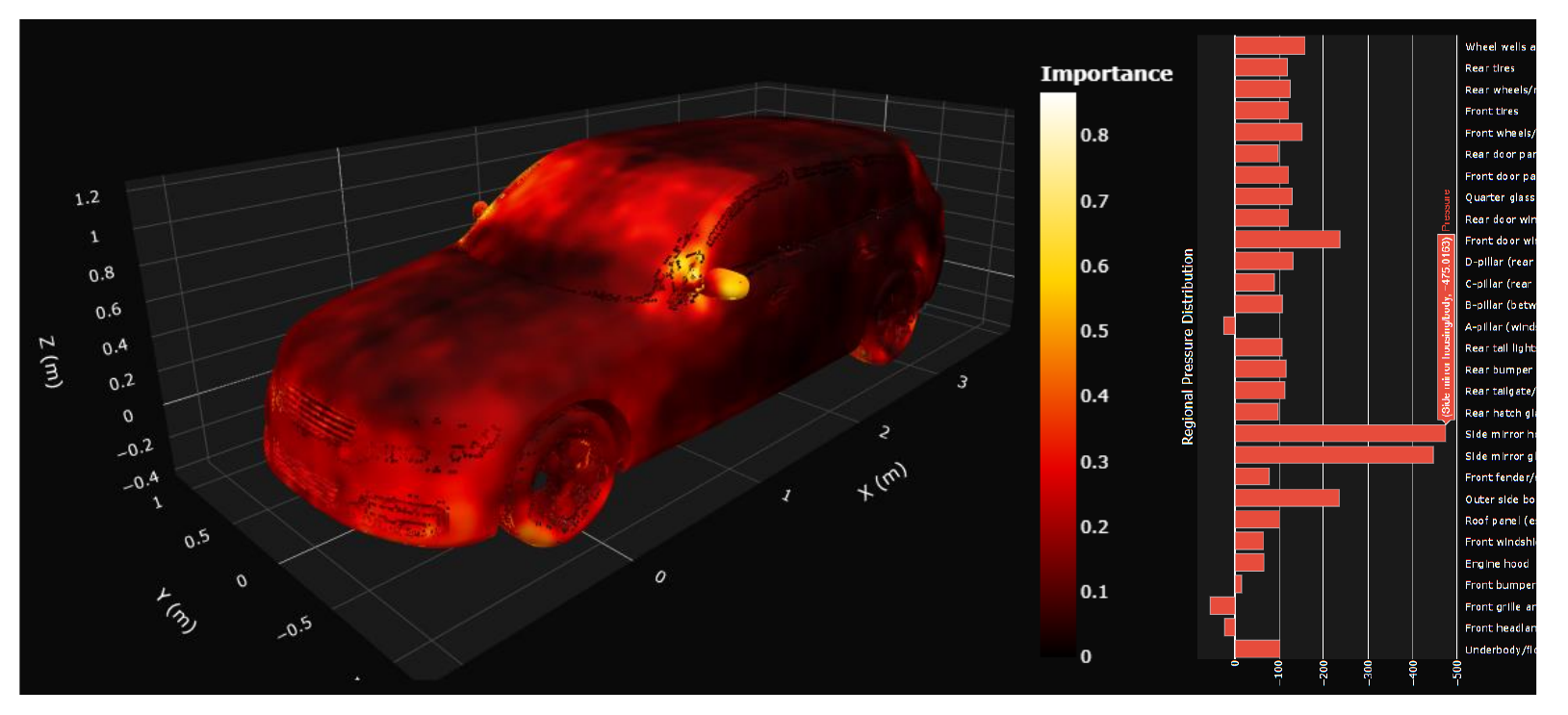}
  \caption{Explainability example highlighting regions that contribute
  to aerodynamic surface field prediction.}
  \label{fig:uc2_xai}
\end{figure}

The same information can also support uncertainty guided data generation.
If predictive uncertainty or model disagreement rises in specific flow regions
or for particular body shape variants, those signals can help engineers decide
which additional CFD cases should be run to expand the training set most
effectively.
This is especially relevant for extending coverage beyond the currently
validated training domain, where new geometries or operating conditions may
expose gaps in the surrogate.

\subsection{Generalizability}

This use case shows that the graph learning philosophy used for BiW mode
classification can also support high-dimensional aerodynamic field prediction.
The shared element is not a single cross-domain pretrained model, but a common
engineering strategy: convert irregular 3D assets into interpretable graphs,
embed domain knowledge through graph construction and supervision, and adapt the
output head to the task.
Within the aerodynamic domain, this strategy can be reused across different
3D body shape variants through a consistent surface graph representation and a
symmetry-aware preprocessing pipeline.

The aerodynamic study nevertheless has a clear scope.
It is restricted to steady state external aerodynamics on the DrivAer dataset family,
uses a surface based surrogate rather than a full 3D flow volume model, and
shows reduced accuracy under more extreme out-of-distribution conditions such as
large yaw angles or substantial spoiler variations.
These limitations do not diminish the present use case; instead, they define
its valid operating range and highlight trustworthy, where explainability and
uncertainty aware analysis are most valuable.
Within a well characterized domain, the AI engineering can provide not only
efficient aerodynamic field prediction, but also interpretable insight into the
flow regions that govern the prediction and practical guidance for targeted data
generation when additional CFD simulations are required.
Future work may build on this foundation through uncertainty guided sampling and
extension to latent-based representation sampling, as well as broader geometric and boundary condition variations.

%% file: conclusion.tex

\section{Discussion and Conclusion}

\begin{figure}[tb]
  \centering
  \includegraphics[width=\columnwidth]{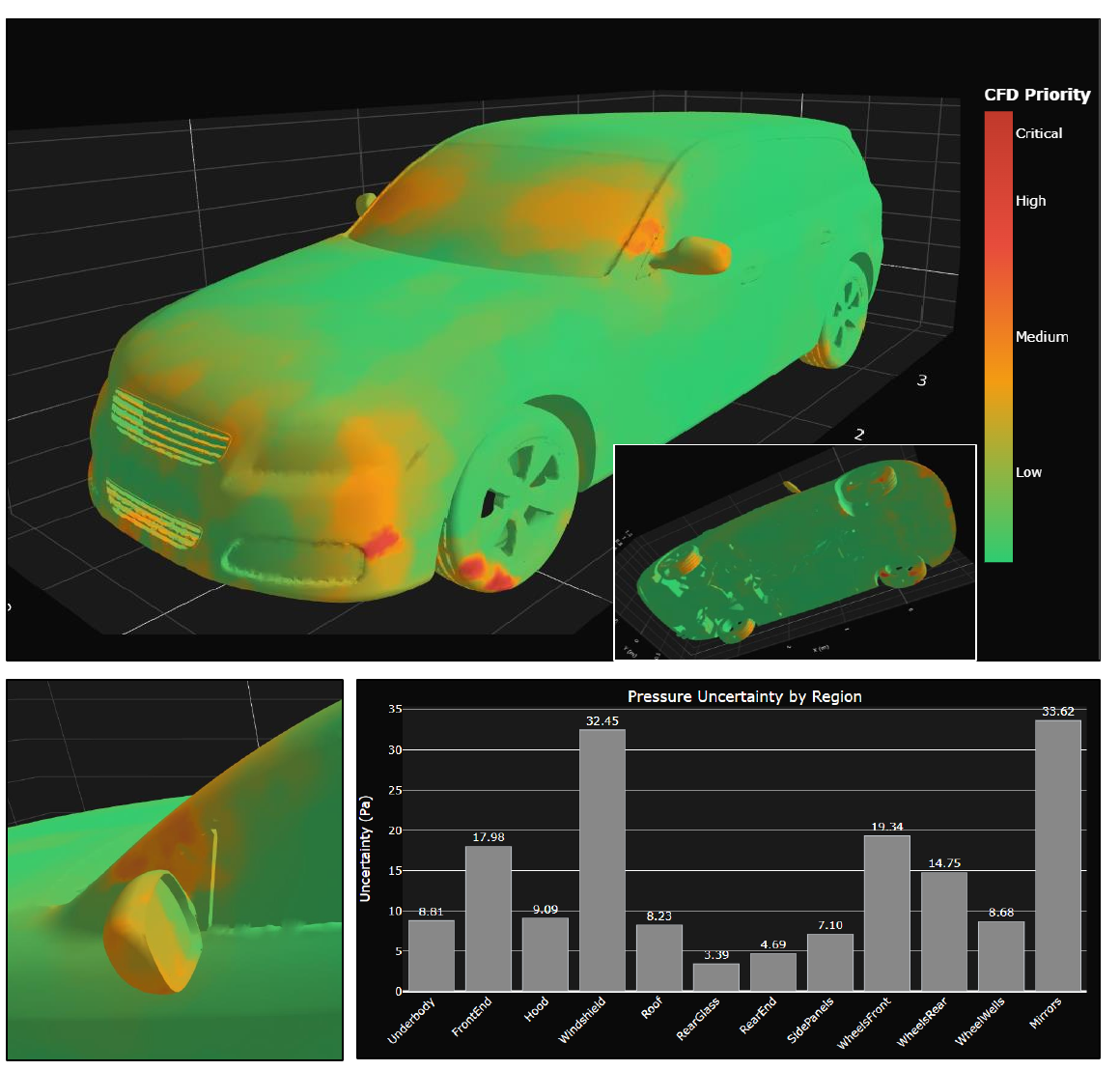}
  \caption{Uncertainty-guided data generation concept. The current framework
  can indicate where additional CFD samples or simulations are likely to be
  most valuable.}
  \label{fig:uc2_activelearning}
\end{figure}

This paper presented a reusable graph learning framework for 3D engineering
AI, in which heterogeneous engineering data are represented as physics-aware
graphs and processed through engineering-aligned workflows.
Across the two automotive use cases, the unifying principle is not a single
cross-domain model, but the combination of engineering-informed graph
representations, task specific prediction, and explanations grounded in
meaningful physical entities.
The contribution is therefore best understood not as a universal foundation
model, but as a practical and reusable engineering AI paradigm for related
tasks and simulation variants.

The results demonstrate that graph-based learning can support different
engineering tasks while preserving domain relevance and interpretability.
In the CAE setting, a canonical regional graph enables mode classification
across BiW and FE variants.
In the CFD setting, a physics-informed surface graph enables pressure and
wall shear stress prediction across body shape variants.
Taken together, these use cases indicate that graph abstraction can provide a
coherent basis for both reuse and interpretation.
From an engineering perspective, the framework offers the potential to reduce
repeated expert effort in structural assessment and repeated field evaluation
in aerodynamic design studies, while enabling earlier feedback and more
effective use of historical engineering data. 

The present work nevertheless has limitations.
The framework is demonstrated on two automotive use cases rather than the
broader range of CAE and CFD applications, and graph construction still
depends on domain knowledge when defining regions, relations, and meaningful
features for new modalities. Overall, this work should be viewed as a step toward trustworthy and reusable
engineering AI workflows based on graph representations, task flexibility, and engineering grounded interpretability.

Future work will focus on broader use cases, data generation methodologies, and furthur integration with engineering practice.

%% file: main.bib
@article{vanderauweraer2007structural,
  title   = {{Virtual Engineering at Work: The Challenges for Designing Mechatronic Products}},
  author  = {Van der Auweraer, Herman and Anthonis, Jan and De Bruyne, Stijn and Leuridan, Jan},
  journal = {Engineering with Computers},
  volume  = {29},
  pages   = {389--408},
  year    = {2013},
  doi     = {10.1007/s00366-012-0286-6}
}

@book{ewins2000modal,
  author    = {Ewins, D. J.},
  title     = {{Modal Testing: Theory, Practice and Application}},
  year      = {2000},
  publisher = {Research Studies Press},
  address   = {London},
  institution = {Imperial College of Science, Technology and Medicine}
}

@article{cfd_automotive,
  title   = {{New Horizons of Vehicle Aerodynamics}},
  author  = {Jessing, Christoph and Stoll, Daniel and Kuthada, Timo and Wiedemann, Jochen},
  journal = {Proceedings of the Institution of Mechanical Engineers, Part D: Journal of Automobile Engineering},
  volume  = {231},
  number  = {9},
  pages   = {1190--1202},
  year    = {2017},
  doi     = {10.1177/0954407017703245}
}

@article{tao2008vehicle,
  title   = {{Mode Calculation and Testing of a Car Body in White}},
  author  = {Yang, Ying and Zhao, Guangyao and Ma, Dongbo and Xu, Xiaobin},
  journal = {Shock and Vibration},
  volume  = {18},
  number  = {1-2},
  pages   = {289--298},
  year    = {2011},
  doi     = {10.3233/SAV-2010-0604}
}

@article{heft2012introduction,
  title   = {{Introduction of a New Realistic Generic Car Model for Aerodynamic Investigations}},
  author  = {Heft, Alexander I. and Indinger, Thomas and Adams, Nikolaus A.},
  journal = {SAE Technical Papers},
  year    = {2012},
  doi     = {10.4271/2012-01-0168}
}

@incollection{gioia2020validation,
  title     = {{Validation of Automatic Modal Parameter Estimator on a Car Body-in-White}},
  author    = {Gioia, N. and Daems, Pieter-Jan and Helsen, J.},
  booktitle = {Topics in Modal Analysis \& Testing, Volume 8},
  pages     = {279--284},
  publisher = {Springer},
  year      = {2020},
  doi       = {10.1007/978-3-030-12684-1_28}
}

@article{bhatnagar2019prediction,
  title   = {{Prediction of Aerodynamic Flow Fields Using Convolutional Neural Networks}},
  author  = {Bhatnagar, Saakaar and Afshar, Yaser and Pan, Shaowu and Duraisamy, Karthik and Kaushik, Shailendra},
  journal = {Computational Mechanics},
  volume  = {64},
  number  = {2},
  pages   = {525--545},
  year    = {2019},
  doi     = {10.1007/s00466-019-01740-0}
}

@inproceedings{drivaernet2024,
  title     = {{DrivAerNet++: A Large-Scale Multimodal Car Dataset with Computational Fluid Dynamics Simulations and Deep Learning Benchmarks}},
  author    = {Elrefaie, Mohamed and Morar, Florin and Dai, Angela and Ahmed, Faez},
  booktitle = {Advances in Neural Information Processing Systems},
  volume    = {37},
  pages     = {499--536},
  year      = {2024},
  note      = {Datasets and Benchmarks Track},
  doi       = {10.52202/079017-0016}
}

@article{drivaerstar2025,
  title   = {{DrivAerStar: An Industrial-Grade CFD Dataset for Vehicle Aerodynamic Optimization}},
  author  = {Qiu, Jiyan and Kuang, Lyulin and Wang, Guan and Xu, Yichen and Cui, Leiyao and Fu, Shaotong and Zhu, Yixin and Zhang, Ruihua},
  journal = {arXiv preprint arXiv:2510.16857},
  year    = {2025},
  doi     = {10.48550/arXiv.2510.16857}
}

@inproceedings{sanchez2020learning,
  title     = {{Learning to Simulate Complex Physics with Graph Networks}},
  author    = {Sanchez-Gonzalez, Alvaro and Godwin, Jonathan and Pfaff, Tobias and Ying, Rex and Leskovec, Jure and Battaglia, Peter W.},
  booktitle = {International Conference on Machine Learning},
  pages     = {8459--8468},
  year      = {2020}
}

@inproceedings{tompson2017accelerating,
  title     = {{Accelerating Eulerian Fluid Simulation With Convolutional Networks}},
  author    = {Tompson, Jonathan and Schlachter, Kristofer and Sprechmann, Pablo and Perlin, Ken},
  booktitle = {International Conference on Machine Learning},
  pages     = {3424--3433},
  year      = {2017}
}

@inproceedings{qi2017pointnet,
  title     = {{PointNet: Deep Learning on Point Sets for 3D Classification and Segmentation}},
  author    = {Qi, Charles R. and Su, Hao and Mo, Kaichun and Guibas, Leonidas J.},
  booktitle = {IEEE Conference on Computer Vision and Pattern Recognition},
  pages     = {652--660},
  year      = {2017}
}

@book{forrester2008engineering,
  title     = {{Engineering Design via Surrogate Modelling: A Practical Guide}},
  author    = {Forrester, Alexander and Sobester, Andras and Keane, Andy},
  publisher = {John Wiley \& Sons},
  year      = {2008}
}

@article{raissi2019physics,
  title   = {{Physics-Informed Neural Networks: A Deep Learning Framework for Solving Forward and Inverse Problems Involving Nonlinear Partial Differential Equations}},
  author  = {Raissi, Maziar and Perdikaris, Paris and Karniadakis, George E.},
  journal = {Journal of Computational Physics},
  volume  = {378},
  pages   = {686--707},
  year    = {2019},
  doi     = {10.1016/j.jcp.2018.10.045}
}

@article{battaglia2018relational,
  title   = {{Relational Inductive Biases, Deep Learning, and Graph Networks}},
  author  = {Battaglia, Peter W. and Hamrick, Jessica B. and Bapst, Victor and Sanchez-Gonzalez, Alvaro and Zambaldi, Vinicius and Malinowski, Mateusz and Tacchetti, Andrea and Raposo, David and Santoro, Adam and Faulkner, Ryan and others},
  journal = {arXiv preprint arXiv:1806.01261},
  year    = {2018}
}

@inproceedings{velickovic2018graph,
  title     = {{Graph Attention Networks}},
  author    = {Veli{\v{c}}kovi{\'c}, Petar and Cucurull, Guillem and Casanova, Arantxa and Romero, Adriana and Li{\`o}, Pietro and Bengio, Yoshua},
  booktitle = {International Conference on Learning Representations},
  year      = {2018}
}

@article{choi2025foundationcs,
  title   = {{Defining Foundation Models for Computational Science: A Call for Clarity and Rigor}},
  author  = {Choi, Youngsoo and Cheung, Siu Wun and Kim, Youngkyu and Tsai, Ping-Hsuan and Diaz, Alejandro N. and Zanardi, Ivan and Chung, Seung Whan and Copeland, Dylan Matthew and Kendrick, Coleman and Anderson, William and Iliescu, Traian and Heinkenschloss, Matthias},
  journal = {arXiv preprint arXiv:2505.22904},
  year    = {2025},
  doi     = {10.48550/arXiv.2505.22904}
}

@article{tohmuang2025modegcn,
  title   = {{Structure Mode Shapes Classification Using Graph Convolutional Networks in Automotive Application}},
  author  = {Tohmuang, Sitthichart and Fard, Mohammad and Marzocca, Pier and Swayze, James L. and Huber, John E. and Fayek, Haytham M.},
  journal = {Computers \& Structures},
  volume  = {314},
  pages   = {107767},
  year    = {2025},
  doi     = {10.1016/j.compstruc.2025.107767}
}

@article{Millan10112025,
author = {Pedro Millan and Benjamin Desai and Tim Cowlam and Eduardo Marques and Lucas F. M. da Silva and Jorge Ambrósio},
title = {{Study of the Effect of Body-in-White Structural Design in Road Vehicle Dynamics}},
journal = {Vehicle System Dynamics},
volume = {0},
number = {0},
pages = {1--25},
year = {2025},
publisher = {Taylor \& Francis},
doi = {10.1080/00423114.2025.2585863}
}

@article{carbench2025,
  title   = {{CarBench: A Comprehensive Benchmark for Neural Surrogates on High Fidelity 3D Car Aerodynamics}},
  author  = {Elrefaie, Mohamed and Shu, Dule and Klenk, Matt and Ahmed, Faez},
  journal = {arXiv preprint arXiv:2512.07847},
  year    = {2025},
  doi     = {10.48550/arXiv.2512.07847}
}

@inproceedings{gal2016dropout,
  title     = {{Dropout as a Bayesian Approximation: Representing Model Uncertainty in Deep Learning}},
  author    = {Gal, Yarin and Ghahramani, Zoubin},
  booktitle = {International Conference on Machine Learning},
  pages     = {1050--1059},
  year      = {2016}
}
